\documentstyle[prl,aps,psfig]{revtex}

\begin{document}

\twocolumn[\hsize\textwidth\columnwidth\hsize\csname@twocolumnfalse\endcsname

\title{The energetics of oxide surfaces by quantum Monte Carlo}

\author{D. Alf\`{e}$^{1,2,3}$, and M. J. Gillan$^{2,3}$
\smallskip \\ 
$^1$Department of Earth Sciences, University College
London \\ Gower Street, London WC1E~6BT, UK
\smallskip \\ 
$^2$Department of Physics and Astronomy, University
College London \\ Gower Street, London WC1E~6BT, UK 
\smallskip \\
$^3$London Centre for Nanotechnology, University College London \\
Gower Street, London WC1E~6BT, UK}

\maketitle

\begin{abstract}
Density functional theory is widely used in surface science, but gives
poor accuracy for surface energetics in many cases. We propose a
practical strategy for using quantum Monte Carlo techniques to correct
DFT predictions, and we demonstrate the operation of this strategy
for the formation energy of the MgO~(001) surface and the adsorption
energy of the H$_2$O molecule on this surface. We note the possibility
of applying the strategy to other surface problems that may be
affected by large DFT errors.
\end{abstract}


]

For many years,  
electronic-structure techniques 
have played a major role in surface and interface 
science. The most widely used of 
these techniques is density functional theory (DFT)~\cite{martin04a}, 
which has been employed to 
study many problems, including 
metal-oxide adhesion, catalysis and corrosion. Yet there is 
evidence that commonly used 
DFT approximations are often seriously in 
error for basic quantities like
surface formation energies and molecular adsorption energies. 
It has been 
noted~\cite{filippi02a} that quantum Monte Carlo (QMC) 
techniques may be able to overcome these 
problems, because of their higher 
accuracy. We propose here
a general strategy for using QMC 
to assess and correct DFT predictions for 
surface formation and molecular 
adsorption energies, based on the idea
that the important DFT errors are 
localised near the surface, so that the 
comparison of QMC with DFT for 
small systems may often suffice to 
give the information needed.
We will show the practical operation of 
this strategy for the formation energy of 
the MgO (001) surface and the adsorption energy of 
water on this surface, comparing where possible
with experimental data.

Insight into DFT errors for surface 
energetics comes from work on the jellium 
surface~\cite{jellium_surface,almeida02a}. Jellium is the homogeneous 
interacting electron gas neutralised by a uniform  
background; its density $n$ is 
characterised by the mean inter-electron 
distance $r_s$, defined by $( 4 \pi r_s^3 / 3 
) n = 1$, with $r_s$ in atomic units. 
The planar jellium surface 
is formed by having the neutralising background 
occupy only the half-space $x < 0$, 
so that the electron number density $n ( x )$ in the ground state
goes to its bulk value $n$ for $x \rightarrow - \infty$ 
and to 0 for $x \rightarrow \infty$.  
Accurate results for the formation energy
of the jellium surface have been obtained~\cite{almeida02a} by 
extrapolating QMC calculations on 
neutral jellium spheres of different 
radius $R$. The extrapolation was performed by 
studying the large-$R$ behaviour of the difference 
between the QMC total energy 
and the total energy calculated with 
DFT approximations, the main such 
approximations being: the local density 
approximation (LDA)~\cite{martin04a}; the generalised 
gradient approximation (GGA) in the 
Perdew-Burke-Ernzerhof form (PBE)~\cite{perdew96a}; 
and the meta-GGA~\cite{meta_GGA}. 
This jellium surface work showed that: 
(i)~for $2 \le r_s 
\le 5$, typical of simple metals 
such as Al and Na, the meta-GGA gives a very 
accurate surface energy $\sigma$, followed closely 
by LDA, with GGA being too low by a significant amount;
(ii)~as $r_s$ falls below 2, 
the GGA errors rapidly worsen. In the region $r_s \sim 1.5$, 
characteristic of transition metals and many 
oxides, the GGA $\sigma$ is 
too low by $\sim 0.3$~J~m$^{-2}$, a serious error,
because the surface energies of 
these materials are themselves in the region 
of $1$~J~m$^{-2}$.

Indirect confirmation for large GGA errors in 
surface energies comes from 
a study~\cite{mattsson02a} of the work of adhesion $W_{\rm adh}$ 
of Pd~(111) to $\alpha$-Al$_2$O$_3$~(0001), 
for which accurate measurements are available. (Here, $W_{\rm adh}$ 
is the reversible work per unit area needed to separate the
system containing the oxide-metal interface into its
metal and oxide consituents.)
The GGA value $W_{\rm adh} = 1.6$~J~m$^{-2}$ 
is far below the LDA 
and experimental values of 2.4 and 2.8~J~m$^{-2}$. 
The authors argue~\cite{mattsson02a} that the 
GGA error comes mainly from errors in the free surface energies, 
and semiquantitatively relate these
errors to GGA errors for the 
jellium surface. They suggest  
the general use of QMC  
jellium surface energies 
to correct DFT predictions for the 
surface energies of real materials~\cite{mattsson02a}.

The strategy we propose also uses QMC to 
correct DFT, but we apply QMC directly 
to the system of interest. In principle, QMC
could be applied by brute force to the large 
slab systems commonly used to model 
surfaces in DFT calculations. However, 
since QMC is far more
costly than DFT, this is not generally feasible 
at present. It is also unnecessary, and not the best way
of gaining insight. 
Since DFT errors for surface energetics 
are expected to be 
localised in the surface region~\cite{mattsson01a}, an accurate 
assessment of these errors should be 
given by QMC calculations only on the 
atoms in the surface region. This implies that 
thin slabs, containing only 
a few atomic layers, should suffice to 
assess the {\em difference} between the 
surface energy given by QMC and by DFT 
approximations, so that this 
difference will converge more rapidly with 
increasing slab thickness than the 
separate surface energies.  
This strategy resembles that 
used to extract the energy of the jellium 
surface from calculations on jellium spheres~\cite{almeida02a}. 
We propose to use the same scheme 
for molecular adsorption energies. The 
molecule is placed on the surface of a thin 
slab, and we study the difference between the 
QMC and DFT adsorption energies, 
seeking convergence of this difference 
with increasing slab thickness. 

We have studied the practical feasibility of this 
strategy for the formation energy 
$\sigma$ of the MgO (001) surface. Our DFT calculations
used the standard pseudopotential/plane-wave 
techniques~\cite{martin04a}, and were
performed using the VASP code~\cite{VASP}. The surface was 
modelled using periodically repeated 
slab geometry, the calculation conditions being 
characterised by basis-set completeness 
(plane-wave cut-off energy 
$E_{\rm cut}$), Brillouin-zone sampling of the electronic states, the 
width $L$ of the vacuum layer separating 
successive slabs, and the number of 
layers $N_{\rm layer}$ in each slab. 
The surface formation energy is $\sigma = 
( E_{\rm slab} - E_{\rm bulk} ) / A$, 
with $E_{\rm slab}$ the energy of the slab 
system, per repeating cell, $E_{\rm bulk}$ the 
energy of the same number of 
atoms of the bulk material, and $A$ the 
total surface area (both faces) of the slab, 
per repeating cell. This definition applies 
for all $N_{\rm layer} \ge 1$. 
The bulk energy $E_{\rm bulk}$ is $N_{\rm layer}$ times the bulk
energy per layer $e_{\rm bulk}$, and it is 
convenient to obtain $e_{\rm bulk}$ from the difference
of $E_{\rm slab}$ values for successive values of $N_{\rm layer}$ in
the limit of large $N_{\rm layer}$.
For given $N_{\rm layer}$, we always insist 
on convergence of the calculated $\sigma$ with respect to $E_{\rm cut}$, 
BZ sampling and $L$ to within 
0.01~J~m$^{-2}$ (this tolerance is satisfied for 
$L > 6$~\AA).  
As expected from earlier work, $\sigma$ for MgO~(001)
converges rapidly with respect to $N_{\rm layer}$,
the residual errors being below 0.01~J~m$^{-2}$ for 
$N_{\rm layer} \ge 2$.
For a given DFT approximation, the calculated 
$\sigma$ depends a little on 
MgO lattice parameter $a_0$. For 
the experimental value $a_0 = 4.21$~\AA, we 
obtain $\sigma = 1.24$ and 0.87~J~m$^{-2}$ 
with LDA and GGA(PBE) respectively. The 
difference of 0.37~J~m$^{-2}$ between 
the two is very similar to the difference of $\sim 0.4$~J~m$^{-2}$
between the LDA and GGA surface energies of 
$\alpha$-Al$_2$O$_3$~(0001)~\cite{mattsson02a}.

The calculation of $\sigma$ by QMC is not standard,
and we are not aware of previous calculations of $\sigma$ for any 
oxide surface using QMC, though our recent
QMC calculations on perfect and 
defective MgO crystals~\cite{alfe05a} indicated
the feasibility of the present calculations. We refer the reader
to reviews for details of QMC (e.g.~\cite{foulkes01a}). 
We recall that for high-precision
results it is essential to use diffusion Monte Carlo (DMC), in which
the many-electron wavefunction is evolved in imaginary time, 
starting from an optimised trial wavefunction generated in prior
variational Monte Carlo calculations. The only error inherent in
DMC is ``fixed-node'' error, due to the fact the nodes of the many-electron
wavefunction are constrained to be those of the trial wavefunction.
For many systems, including jellium, the evidence is that fixed-node error
is extremely small. For wide-gap systems such as MgO, the errors should
be no greater. Our calculations were performed with the
{\sc casino} code~\cite{needs04a}, using the same Hartree-Fock pseudopotentials
as in our previous work~\cite{alfe05a}. The trial wavefunctions
were of the usual Slater-Jastrow type, with single-electron
orbitals obtained with the plane-wave code PWSCF~\cite{pwscf}, 
generally using the large plane-wave cut-off of 4082~eV. These orbitals
were represented in {\sc casino} using the recently
reported ``blip-function'' real-space basis set~\cite{alfe04a}.
The DMC calculations all used a time-step of 0.005~a.u., and mean number of
walkers equal to 10,240. The calculations were done with free
boundary conditions (i.e. no periodicity) normal to the surface.

Our DMC calculations were performed on a series of MgO slabs with 
the number of layers $N_{\rm layer}$ running from 1 to 5.
For each $N_{\rm layer}$, convergence must be demonstrated with respect
to basis-set completeness and size of repeating surface unit cell. Basis-set
errors with the blip basis set are readily made negligible, as shown
earlier~\cite{alfe04a}. In DMC calculations, the wavefunctions
are real, so that Brillouin-zone sampling is generally impossible,
and calculations are usually performed at the $\Gamma$-point; this is
why convergence with respect to size of surface unit cell must be checked.
Our main DMC calculations used the $2 \times 2$ surface unit cell, for which
the repeating cells contain from 16 ($N_{\rm layer} = 1$) to
80 ($N_{\rm layer} = 5$) ions; we show below that larger surface cells
would give almost identical results. 

The raw output from these
calculations is DMC total energies $E_{\rm slab}^{\rm DMC}$ for
the five $N_{\rm layer}$ values. Following our strategy,
we now study the difference $\Delta E_{\rm slab} \equiv
E_{\rm slab}^{\rm DMC} - E_{\rm slab}^{\rm DFT}$, with the DFT
slab energy calculated with exactly the same slab and 
the same ($\Gamma$-point) BZ sampling as in the DMC calculations.
Since the jellium results indicate that LDA surface energies are likely
to be closer to DMC than those from GGA, we use LDA values for
$E_{\rm slab}^{\rm DFT}$. When plotted against $N_{\rm layer}$, 
$\Delta E_{\rm slab}$ will tend asymptotically 
($N_{\rm layer} \rightarrow \infty$) to a straight line, whose slope
is equal to the difference between the DMC and LDA bulk
energies $\Delta e_{\rm bulk}$ per layer, and whose $N_{\rm layer} = 0$
intercept divided by $A$ gives the difference of DMC and DFT surface
energies $\Delta \sigma \equiv \sigma^{\rm DMC} - \sigma^{\rm DFT}$.
Since $\Delta e_{\rm bulk}$ is large, and since $\Delta E_{\rm slab}$ contains
the statistical errors of DMC, it is helpful to start this analysis
by performing a least-squares straight-line fit
$a + b N_{\rm layer}$ to the values of $\Delta E_{\rm slab}$,
and then to use the resulting $b$ value to form the
quantity $\tilde{\Delta} E_{\rm slab} \equiv
\Delta E_{\rm slab} - b N_{\rm layer}$.
The $N_{\rm layer} \rightarrow \infty$ straight-line asymptote
of $\tilde{\Delta} E_{\rm slab}$ has the same $N_{\rm layer} = 0$
intercept as that of $\Delta E_{\rm slab}$. For
$N_{\rm layer} = 1, 2, \ldots 5$, we find the five
values $\tilde{\Delta} E_{\rm slab} = -0.019(2)$, $-0.009(6)$, $-0.007(9)$,
$-0.011(13)$ and $-0.014(15)$~J~m$^{-2}$. This immediately shows that
the DMC and LDA values of $\sigma$ are almost exactly the same. Within our
rather small statistical errors of at worst
0.015~J~m$^{-2}$, the difference between the
DMC and LDA surface energies has the very small
value of $-0.01$~J~m$^{-2}$. 
To check the errors due to use of the $2 \times 2$ surface cell (i.e. 
errors of BZ sampling), we have performed LDA calculations
on slabs having large surface cells with a series of $N_{\rm layer}$
values, using $\Gamma$-point sampling. The $\sigma$ values thus
obtained differ from the LDA $\sigma$ value extracted 
by similar $\Gamma$-point calculations with the $2 \times 2$ surface cell
by only 0.01~J~m$^{-2}$.
The LDA value of $\sigma$ for the lattice parameter we are using,
converged with respect to BZ sampling and slab thickness,
is 1.20~J~m$^{-2}$, and we conclude from our 
$\tilde{\Delta} E_{\rm slab}$ values that the fully converged
DMC value is $1.19 \pm 0.01$~J~m$^{-2}$.
MgO is one of the few oxides for which reasonably reliable experimental
values of the surface energy are available~\cite{tosi64a}. Exploiting the fact
that MgO cleaves readily along the (001) plane, the experiments
measure the work of cleavage, thus
ensuring that the results for $\sigma$ cannot be influenced
by surface contamination. Our $\sigma_{\rm DMC}$ value of
1.19~J~m$^{-2}$ is consistent with the measured 
values~\cite{tosi64a}, which fall
in the range $1.04 - 1.20$~J~m$^{-2}$.

\begin{figure}
\psfig{figure=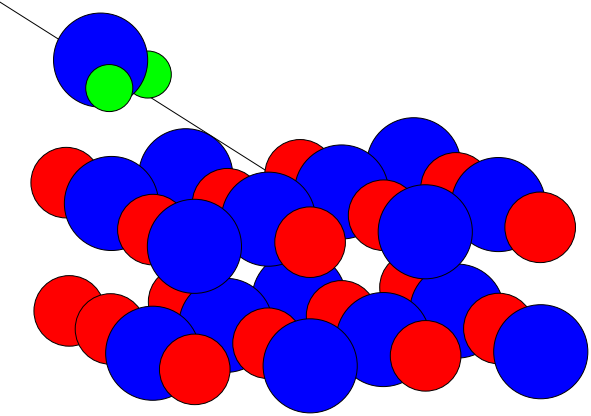,height=2.2in,angle=-90}
\caption{Adsorption geometry of the H$_2$O molecule on MgO(001) from
DFT with PBE exchange-correlation functional.}\label{fig:h2o_mgo}
\end{figure}

In applying QMC to correct DFT predictions for the adsorption energy
$E_{\rm ads}$ of H$_2$O on MgO~(001), we assume, in accord with
experimental and theoretical indications~\cite{xu97a,scamehorn93a}, 
that the molecule lies
almost flat on the surface, with the water O atom almost above a surface
Mg ion, the water O--H bonds pointing towards 
surface O ions (Fig.~\ref{fig:h2o_mgo}). The adsorption
energy is defined to be $E_{\rm ads} = E_{{\rm H}_2 {\rm O}} +
E_{{\rm bare} \; {\rm slab}} - E_{{\rm slab} + {\rm H}_2 {\rm O}}$,
where the terms on the right are the energy of the isolated H$_2$O molecule,
the energy of the bare MgO~(001) slab, and the energy of the slab with
the H$_2$O molecule adsorbed on the surface, all three systems being
fully relaxed to equilibrium. In our DFT calculations, we require that
$E_{\rm ads}$ be converged to within 10~meV with respect to plane-wave
cut-off $E_{\rm cut}$, BZ sampling, and vacuum width $L$. Furthermore,
since we want $E_{\rm ads}$ for an isolated molecule, we examine the
dependence of $E_{\rm ads}$ on the size of the surface unit cell. We find
that with the $2 \times 2$ cell, $E_{\rm ads}$ is already converged
to better than 10 meV (this was tested by doing calculations 
up to $5 \times 5$ surface unit cells). With 
these tolerances always applied, we then study the
dependence of $E_{\rm ads}$ on the number of layers in the slab
$N_{\rm layer}$. As $N_{\rm layer}$ increases, $E_{\rm ads}$ ceases 
to change by more than 2~meV for $N_{\rm layer} \ge 2$. Our
calculated values of $E_{\rm ads}$ with LDA and GGA(PBE)
are 0.92 and 0.43~eV respectively.

In the DMC calculations, we obtained $E_{{\rm H}_2 {\rm O}}$ using
periodically repeated cubes of different lengths $d$, with a single
molecule in each cube, using the molecular geometry taken from
PBE (O--H bond length = 0.978~\AA, bond-angle
= 104.4$^\circ$). There is a 
weak dipole-dipole correction, going as
$d^{-3}$, but we extrapolate to infinite $d$ to obtain $E_{{\rm H}_2 {\rm O}}$
with a technical error uncertainty of only a few meV. The slab energy
$E_{{\rm bare} \; {\rm slab}}$ is taken from our DMC calculations
on $\sigma$ (see above). For H$_2$O on the slab, strict application
of our strategy would require us to use the relaxed configuration
obtained from DMC calculations. We are not yet able to do this,
since the calculation of ionic forces with DMC is problematic for
the moment (though see Ref.~\cite{lee05a}). Instead, we use a relaxed
configuration from the DFT approximation which appears to reproduce DMC
most closely. Our DFT calculations show that the relaxed height of
the molecule above the surface differs by $\sim 0.15$~\AA\ between
LDA and PBE. We calculated the DMC energy fo H$_2$O on the 1-layer slab,
with a series of geometries on a linear path between the relaxed LDA and
PBE geometries, and we find that the PBE geometry is very close to
giving the lowest DMC energy. All our DMC results therefore
refer to relaxed PBE geometries.

Since DFT calculations of $E_{\rm ads}$ are converged with respect to
surface cell size for $2 \times 2$ cells, our DMC calculations
are all done with this surface cell. With $N_{\rm layer} = 1$,
we find $E_{\rm DMC}^{\rm ads} = 0.63(3)$~eV, $E_{\rm PBE}^{\rm ads} = 0.48$~eV,
so that $\Delta E^{\rm ads} \equiv E_{\rm DMC}^{\rm ads} - 
E_{\rm PBE}^{\rm ads} = 0.15$~eV. For $N_{\rm layer} = 2$,
the results are $E_{\rm DMC}^{\rm ads} = 0.57(4)$~eV, 
$E_{\rm PBE}^{\rm ads} = 0.42$~eV, so that
$\Delta E^{\rm ads} = 0.15$~eV, identical to the $N_{\rm layer} =1$
value within statistical errors. Using the $N_{\rm layer} \rightarrow \infty$
PBE value of 0.43~eV, we thus estimate the $N_{\rm layer} \rightarrow \infty$
DMC value as 0.58(3)~eV.
 
For $E_{\rm ads}$, a comparison with experiment can only be indicative
at present. Measurements of LEED isotherms and isobars for
H$_2$O adsorption on MgO~(001) as a function of coverage, extrapolated
to zero coverage gives $E_{\rm ads} = 0.52 \pm 0.10$~eV~\cite{ferry98a}. 
Temperature programmed desorption experiments show a peak at $T = 235 - 260$~K
due to desorption of water at initial monolayer 
coverage~\cite{xu97a,stirniman96a}. The standard
Redhead analysis, using the commonly assumed frequency prefactor
of $10^{13}$~sec$^{-1}$, yields an effective $E_{\rm ads}$ of
$0.63 - 0.67$~eV. However, this must include a significant
contribution from attractive water-water interactions, so that
$E_{\rm ads}$ for isolated H$_2$O should be somewhat lower, and thus perhaps
consistent with the LEED value of $0.52 \pm 0.10$~eV. To compare our
DMC value with this, a correction for vibrational energies is needed.
We have performed GGA(PBE) calculations on the adsorbed molecule,
which indicate that adsorption lowers the symmetric and asymmetric
stretch modes of H$_2$O by 12 and 9~THz respectively, and raises the
bond-bending mode by 2~THz. The associated zero-point energies raise
the adsorption energy by 39~meV. Translational and rotational
energies of the H$_2$O molecule in free space, and vibrations
and librations of the adsorbed molecule relative 
to the surface~\cite{vibrations}
raise and lower the adsorption energy by 64~meV and 167~meV
respectively. Altogether, vibrational effects lower the adsorption
energy by 64~meV, so that our corrected DMC adsorption energy
is 0.51~eV. Our calculated value thus appears to be consistent
with the experimental evidence, but clearly a more elaborate analysis
would be needed to make this comparison robust. Our comparisons
make it fairly clear that LDA gives a serious overestimate of the
adsorption energy, while GGA (PBE) is considerably more accurate.

Our proposed strategy is therefore feasible and useful for MgO
surface energetics. Many other important problems could be addressed
in the same way, including the $\alpha$-Al$_2$O$_3$~(0001) surface
energy mentioned earlier~\cite{mattsson02a}; 
the computational effort needed for this
case would similar to MgO. For this and other applications of the strategy,
we believe that current progress in the calculation of ionic forces
and structural relaxation with DMC~\cite{lee05a} 
will be very helpful.

The ability to validate QMC against experiment for molecular adsorption
energies is limited by the lack of adequate techniques for putting 
modelling and data into close contact. Progress in the {\em ab initio}
prediction of TPD spectra~\cite{stampfl99a} is encouraging in this
respect. Finally, we note the importance of quantum chemistry methods.
For wide-gap materials, techniques such as MP2 and CCSD(T) should
deliver an accuracy similar to that of DMC. Major improvements in
the scaling of these techniques with number of atoms and size of basis
sets~\cite{werner03a} and to perform them in periodic boundary
conditions~\cite{pisani05a} should make it possible to use them within
our proposed strategy. 

In summary, we have described a practical general strategy for
using QMC calculations to assess and correct the errors of DFT
approximations for the energetics of surfaces. Our calculations
on the surface formation energy of MgO~(001) and the adsorption energy
of H$_2$O on this surface confirm the feasibility and usefulness
of the strategy. The results support earlier inferences from
the energetics of the jellium surface that the GGA surface formation
energy for this type of material is substantially too low, and that
LDA is more accurate. However, for the molecular adsorption energy,
the reverse is true, with LDA errors being much greater than those
of GGA.

DA acknowledges support from the Royal Society and EPSRC. Allocation of 
computer time at the HPCx national service was provided by the Materials
Chemistry Consortium, the UKCP Consortium and the Mineral Physics
Consortium. We thank the Texas Advanced Computing Center at the
University of Texas at Austin for providing high-performance
computing resources. MJG acknowledges useful discussions with F.~Manby
and M.~Scheffler.

\begin{thebibliography}{99}

\bibitem{martin04a}
R. M. Martin, {\em Electronic Structure: Basic Theory and Practical Methods},
Cambridge University Press, Cambridge (2004).

\bibitem{filippi02a}
C. Filippi, S. B. Healy, P. Kratzer, E. Pehlke, and M. Scheffler,
Phys. Rev. Lett. {\bf 89}, 166102 (2002).

\bibitem{jellium_surface}
P. H. Acioli and D. M. Ceperley, Phys. Rev. B {\bf 54}, 17199 (1996);
F. Sottile and P. Ballone, {\em ibid.} {\bf 64}, 045105 (2001);
Z. Yan, J. P. Perdew, S. Kurth, C. Fiolhais, L. Almeida, {\em ibid.} 
{\bf 61}, 2595 (2000); J. M. Pitarke, {\em ibid.} {\bf 70},
087401 (2004).

\bibitem{almeida02a}
L. M. Almeida, J. P. Perdew and C. Fiolhais, Phys. Rev. B {\bf 66},
075115 (2002).

\bibitem{perdew96a}
J. P. Perdew, K. Burke, and M. Ernzerhof, Phys. Rev. Lett {\bf 77},
3865 (1996).

\bibitem{meta_GGA}
J. P. Perdew, S. Kurth, A. Zupan, P. Blaha, Phys. Rev. Lett. {\bf 82},
2544 (1999); J. Tao, J. P. Perdew, V. N. Staroverov, G. E. Scuseria,
{\em ibid.} {\bf 91}, 146401 (2003).

\bibitem{mattsson02a}
A. E. Mattsson and D. R. Jennison, Surf. Sci. Lett. {\bf 520}, L611 (2002).

\bibitem{mattsson01a}
A. E. Mattsson and W. Kohn, J. Chem. Phys. {\bf 115}, 3441 (2001).

\bibitem{VASP}
J. Kresse and J. Furthm\"{u}ller, Phys. Rev. B {\bf 54}, 11169 (1996).

\bibitem{alfe05a} D. Alf\`e, M. Alfredsson, J. Brodholt, M. J. Gillan,
M. D. Towler, and R. J. Needs, Phys. Rev. B, {\bf 72}, 014114 (2005);
D. Alf\`e and M. J. Gillan, {\em ibid.} {\bf 71}, 220101(R) (2005).

\bibitem{foulkes01a} W. M. C. Foulkes, L. Mita\v{s}, R. J. Needs, and
G. Rajagopal, Rev. Mod. Phys. {\bf 73}, 33 (2001).

\bibitem{needs04a} R. J. Needs, M. D. Towler, N. D. Drummond, and
P. R. C. Kent, `{\sc casino} Version 1.7 User Manual', University of
Cambridge, Cambridge (2004).

\bibitem{pwscf} S. Baroni, A. Dal Corso, S. de Gironcoli, and
P. Giannozzi, http://www.pwscf.org.

\bibitem{alfe04a} D. Alf\`e and M. J. Gillan, Phys. Rev. B {\bf 70},
161101(R) (2004).

\bibitem{tosi64a}
M. P. Tosi, in {\em Solid State Physics}, Vol.~16, eds. F.~Seitz and
D. Turnbull (Academic Press, New York, 1964), p.~1 and references therein.

\bibitem{xu97a}
C. Xu and D. W. Goodman, Chem. Phys. Lett. {\bf 265}, 341 (1997).

\bibitem{scamehorn93a}
C. A. Scamehorn, A. C. Hess, and M. I. McCarthy, 
J. Chem. Phys. {\bf 99}, 2786 (1993);
C. A. Scamehorn, N. M. Harrison, and M. I. McCarthy,
J. Chem. Phys. {\bf 101}, 1547 (1994);
O. Engkvist and A. J. Stone, Surf. Sci. {\bf 437}, 239 (1999).

\bibitem{ferry98a}
D. Ferry, S. Picaud, P. N. M. Hoang, C. Girardet, L. Giordano,
B. Demirdjian, and J. Suzanne, Surf. Sci. {\bf 409}, 101 (1998);
D. Ferry, A. Glebov, V. Senz, J. Suzanne, J. P. Toennies, and H. Weiss, 
{\em ibid.} {\bf 377}, 634 (1997).

\bibitem{stirniman96a}
M. J. Stirniman, C. Huang, R. S. Smith, S. A. Joyce, and B. D. Kay, 
J. Chem. Phys. {\bf 105}, 1295 (1996).

\bibitem{vibrations}
The correct calculation of vibrational and librational energies of the
adsorbed molecule in DFT is technically challenging, since these motions
are coupled to the lattice vibrations. For our estimates, we treat the
lattice as rigid. Because of this approximation, the lowering of
$E_{\rm ads}$ by these modes may be in error by $\sim 50$~meV.

\bibitem{lee05a}
M. W. Lee, M. Mella, and A. M. Rappe, J. Chem. Phys. {\bf 122},
244103 (2005).

\bibitem{stampfl99a}
C. Stampfl, H. J. Kreutzer, S. H. Payne, H. Pfn\"{u}r, and M. Scheffler,
Phys. Rev. Lett. {\bf 83}, 2993 (1999).

\bibitem{werner03a}
H.-J. Werner, F. R. Manby, and P. J. Knowles, J. Chem. Phys. {\bf 118},
8149 (2003); M. Sch\"{u}tz, H.-J. Werner, and F. R. Manby, J. Chem. Phys. 
{\bf 121}, 737 (2004).

\bibitem{pisani05a}
C. Pisani, G. Capecchi, S. Cassassa, and L. Maschio, Molec. Phys.
{\bf 103}, 2527 (2005).

\end{thebibliography}
\end{document}